%% file: paper.tex
\documentclass[final]{siamltex}
\usepackage{amsfonts}
\usepackage{amssymb}
\usepackage{amsmath}
\usepackage{stmaryrd}
\usepackage[colorlinks]{hyperref}
\usepackage{graphicx}
\usepackage{algorithm}
\usepackage{psfrag}

\newcommand{\avg}[1]{\langle {#1} \rangle}
\newcommand{\jump}[1]{\llbracket {#1} \rrbracket}

\newcommand{\brac}[1]{\left( {#1} \right)}
\newcommand{\bracc}[1]{\left\{ {#1} \right\}}
\newcommand{\vect}[1]{\boldsymbol{#1} }
\newcommand{\R}{\mathbb{R}}
\newcommand{\dx}{\, \mathrm{d}x}
\newcommand{\dX}{\, \mathrm{d}X}
\newcommand{\ds}{\, \mathrm{d}s}
\newcommand{\tab}{\hspace*{2em}}

\newcommand{\ffc}{FFC}
\newcommand{\dolfin}{DOLFIN}
\newcommand{\fenics}{FEniCS}

\newcommand{\unorm}[1]{%
  \left\vert\kern-0.9pt\left\vert\kern-0.9pt\left\vert #1
    \right\vert\kern-0.9pt\right\vert\kern-0.9pt\right\vert}

\usepackage{verbatim}
\newenvironment{code}[1]%
{\center\tabular{c}\hline\\ \footnotesize\minipage{#1\textwidth}\verbatim}
{\endverbatim\endminipage\\ \\ \hline\endtabular\endcenter}

\title{Automated code generation for discontinuous Galerkin methods}
\author{Kristian B. {\O}lgaard\footnotemark[2]
  \and Anders Logg\footnotemark[3]
  \and Garth N. Wells\footnotemark[4]}

\begin{document}
\maketitle

\renewcommand{\thefootnote}{\fnsymbol{footnote}}
\footnotetext[2]{Faculty of Civil Engineering and Geosciences,
                Delft University of Technology,
                Stevinweg 1, 2628 CN Delft, Netherlands
                ({\tt k.b.oelgaard@tudelft.nl})}
\footnotetext[3]{Center for Biomedical Computing, Simula Research Laboratory /
                 Department of Informatics, University of Oslo,
                 P.O.Box~134, 1325 Lysaker, Norway
                 ({\tt logg@simula.no})}
\footnotetext[4]{Department of Engineering, University of Cambridge, Trumpington Street,
                Cambridge CB2~1PZ, United Kingdom  ({\tt gnw20@cam.ac.uk})}
\renewcommand{\thefootnote}{\arabic{footnote}}
\begin{abstract}
A compiler approach for generating low-level computer code from high-level
input for discontinuous Galerkin finite element forms is presented. The
input language mirrors conventional mathematical notation, and the
compiler generates efficient code in a standard programming language. This
facilitates the rapid generation of efficient code for general equations
in varying spatial dimensions. Key concepts underlying the compiler
approach and the automated generation of computer code are elaborated. The
approach is demonstrated for a range of common problems, including the
Poisson, biharmonic, advection--diffusion and Stokes equations.
\end{abstract}
\section{Introduction}

Discontinuous Galerkin methods in space have emerged as a generalisation
of finite element methods for solving a range of partial differential
equations. While historically used for first-order hyperbolic equations,
discontinuous Galerkin methods are now applied to a range of hyperbolic,
parabolic and elliptic problems. In addition to the usual integration
over cell volumes that characterises the conventional finite element
method, discontinuous Galerkin methods also involve the integration of
flux terms over interior facets. Discontinuous Galerkin methods exist
in many variants, and are generally distinguished by the form of the
flux on facets. A sample of fluxes for elliptic problems can be found
in~\cite{arnold_et_al:2002}.

We present here a compiler approach for generating computer code for
discontinuous Galerkin forms.  From a high-level input language which
resembles conventional mathematical notation, low-level computer code is
generated automatically.  The generated code is called by an assembler
to construct global sparse tensors, commonly known as the `stiffness
matrix' and the `load vector'. The compiler approach affords a number of
interesting possibilities. It permits the rapid prototyping and testing of
new methods, as well as providing scope for producing optimised code. The
latter can be achieved through the compiler by precomputing various
terms which are traditionally evaluated at run time, and by deploying
procedures for analysing the structure of forms which facilitates various
\emph{a priori} optimisations which may not be tractable when developing
computer code in a conventional fashion.  In addition, the representations
of element tensors (element stiffness matrices) for a given variational
form are not limited to the usual quadrature-loop approach.  For many
forms, computationally more efficient representations can be employed.
In essence, the form compiler approach allows a high level of generality,
while competing in terms of performance with specialised, dedicated code,
as will be elaborated in this work.

The use of a form compiler is particularly attractive for mixed
problems, where one may wish to work with a combination of continuous
and discontinuous function spaces, and function spaces which
differ on element interiors.  Such problems become trivial in the
context of a compiler, as the compiler can automatically generate a
degree-of-freedom mapping, thereby alleviating a difficulty faced when
using many legacy codes for mixed problems.  It also bears emphasis
that the compiler provides the necessary operators to generate code not
only for discontinuous Galerkin methods, but also for a range of novel
finite element methods that draw upon discontinuous Galerkin methods.
These methods may not involve discontinuous function spaces but do
involve integration over interior facets. Such examples can be found
in~\cite{hughes:2006,wells:2007,Labeur:2007}.

The concepts presented in this work are implemented in the FEniCS Form
Compiler (henceforth \ffc{}). \ffc{} is a component of the FEniCS
project~\cite{fenics}, which consists of a suite of tools which aim
to automate computational mathematical modelling, and all components
are released under a GNU public license. \ffc{} is freely available
at \url{http://www.fenicsproject.org} and will generate code for all examples
presented in this work.

The rest of this work is arranged as follows.
Section~\ref{sec:compiling_forms} considers aspects of the assembly of
variational forms and the representation of finite element tensors. This
is followed by key concepts for the assembly of discontinuous Galerkin
forms in Section~\ref{sec:dg}. We discuss the form compiler \ffc{} and the
performance of the code it generates in Section~\ref{sec:implementation}
and a number of examples are presented in Section~\ref{sec:examples}.

\section{Compiling and assembling finite element variational forms}
\label{sec:compiling_forms}

In this section, we outline a general framework for compiling
and assembling variational forms. We then extend the framework to
discontinuous Galerkin methods in the following section, where the form
compiler must also consider integrals of discontinuous integrands over
interior facets\footnote{A \emph{facet} is a topological entity of a
  computational mesh of dimension $D-1$ (codimension~$1$) where $D$ is
  the topological dimension of the cells of the computational mesh. Thus
  for a triangular mesh, the facets are the edges and for a tetrahedral
  mesh, the facets are the faces.}
of the computational mesh.
\subsection{Multilinear forms}
Consider a general multilinear form,
\begin{equation}
\label{eq:multilinearform}
  a : V_h^1 \times V_h^2 \times \cdots \times V_h^r \rightarrow \R,
\end{equation}
defined on the product space $V_h^1 \times V_h^2 \times \cdots \times
V_h^r$ of a sequence $\{V_h^j\}_{j=1}^r$ of discrete function spaces on
a triangulation $\mathcal{T}$ of a domain $\Omega \subset \R^d$.

Multilinear forms appear as the basic building blocks of finite element
discretisations of partial differential equations. The canonical example
is the standard variational formulation of Poisson's equation $-\Delta
u = f$. It reads: find $u_h \in V_h$ such that
\begin{equation}
\label{eq:varproblem}
  a(v, u_h) = L(v) \quad \forall \ v \in \hat{V}_h,
\end{equation}
where $\hat{V}_h = V_h^1$ and $V_h = V_h^2$ is a pair of discrete finite
element function spaces (the \emph{test} and \emph{trial} spaces). The
\emph{bilinear} form~$a$ is here given by
\begin{equation}
  a(v, u) = \int_{\Omega} \nabla v \cdot \nabla u \dx
\end{equation}
and the \emph{linear} form~$L$ is given by
\begin{equation}
  L(v) = \int_{\Omega} v f \dx.
\end{equation}
For this problem, $r = 2$ for the bilinear form~$a$ and $r = 1$
for the linear form~$L$, but forms of higher arity also appear
(see~\cite{logg:article:12}).
\subsection{Finite element assembly}

With $\{\phi_i\}_{i=1}^N$ a global basis for $\hat{V}_h = V_h$, one may
obtain the solution $u_h = \sum_{i=1}^N U_i \phi_i$ of the variational
problem~(\ref{eq:varproblem}) by solving a linear system $AU = b$
for the \emph{degrees of freedom} $U$ of the discrete solution~$u_h$,
where $A_{ij} = \int_{\Omega} \nabla \phi_i \cdot \nabla \phi_j
\dx$ and $b_i = \int_{\Omega} \phi_i f \dx$.  In general, we are
concerned with the discretisation of the general multilinear form~$a$
of~(\ref{eq:multilinearform}), that is, the computation of the (sparse)
rank~$r$ tensor~$A$ obtained by applying the multilinear form to the basis
functions $\{\phi_i^j\}_{i=1}^{N_j}$ of $V_h^j$ for $j = 1,2, \ldots, r$:
\begin{equation}
\label{eq:generic_multilinear_form}
  A_i = a(\phi_{i_1}^1, \phi_{i_2}^2, \ldots, \phi_{i_r}^r),
\end{equation}
where $i = (i_1, i_2, \ldots, i_r)$ is a multi-index.

The tensor~$A$ may be computed efficiently by an algorithm known as
\emph{assembly}~\cite{ZieTay67,Hug87,Lan99}. This algorithm computes
the tensor~$A$ by iterating over the elements of the triangulation
$\mathcal{T} = \{K\}$ and adding the local contribution from
each local element~$K$ to the global tensor~$A$. We refer to
the local contribution from each element as the \emph{element
tensor}~$A^K$~\cite{logg:article:12}. For $r = 2$, this is
normally referred to as the `element stiffness matrix'. With
$\{\phi^{K,j}_i\}_{i=1}^{n_j}$ a local basis for $V_h^j$ on element~$K$,
the element tensor~$A^K$ is given by
\begin{equation}
  A^K_i =
  a_K(\phi_{i_1}^{K,1}, \phi_{i_2}^{K,2}, \ldots, \phi_{i_r}^{K,r}),
\end{equation}
where $a_K$ is the local contribution to the multilinear form from
element~$K$. In the case of the bilinear form for Poisson's equation,
this contribution is given by $a_K(v, u) = \int_{K} \nabla v \cdot \nabla
u \dx$.
\subsection{Tensor representation}
In~\cite{logg:article:07,logg:article:09,logg:article:10,logg:article:11},
it was demonstrated that by generating low-level code from a special
tensor representation of the element tensor~$A^K$, one may generate
(compile) very efficient code for assembly of the corresponding global
matrix~$A$. We return to the issue of the efficiency of the generated
code in Section~\ref{sec:performance}.

We demonstrate here how to derive this tensor representation for a
basic example below and refer to~\cite{logg:article:11} for a general
representation theorem.  Consider the bilinear form for the weighted
Laplacian $-\nabla \cdot (w \nabla u) = f$,
\begin{equation}
\label{eq:weightedlaplacian}
  a(v, u) = \int_{\Omega} w \nabla v \cdot \nabla u \dx.
\end{equation}
The corresponding element tensor~$A^K$ is given by
\begin{equation}
  A^K_i = \int_{K} w \nabla \phi^K_{i_1} \cdot \phi^K_{i_2} \dx.
\end{equation}
For simplicity, we consider here the case where $V_h^1 = V_h^2 = V_h$.
Now, let $F_K : K_0 \rightarrow K$ be the standard affine mapping from
a reference element~$K_0$ to any given element $K \in \mathcal{T}$.
Using a change of variables from the reference coordinates $X$ to the
real coordinates $x = F_K(X)$, we find that
\begin{equation}
  A^K_i = \sum_{\alpha_1=1}^d \sum_{\alpha_2=1}^d \sum_{\alpha_3=1}^n
  \det F_K' w_{\alpha_3}
  \sum_{\beta=1}^d
  \frac{\partial X_{\alpha_1}}{\partial x_{\beta}}
  \frac{\partial X_{\alpha_2}}{\partial x_{\beta}}
  \int_{K_0} \Phi_{\alpha_3}
  \frac{\partial \Phi_{i_1}}{\partial X_{\alpha_1}}
  \frac{\partial \Phi_{i_2}}{\partial X_{\alpha_2}}
  \dX,
\label{eq:weightedlaplacian,tensorrepresentation}
\end{equation}
where $\Phi$ denotes basis functions on the reference element, $n$
is the number of degrees of freedom for the local basis of~$w$, $d$
is the dimension of the domain $\Omega$, and we have expanded~$w$ in
the local nodal basis of $V_h$.  By defining the two tensors
$A^0 = \int_{K_0} \Phi_{\alpha_3}
\frac{\partial \Phi_{i_1}}{\partial X_{\alpha_1}}
\frac{\partial \Phi_{i_2}}{\partial X_{\alpha_2}}
\dX$
and
$G_K = \det F_K' w_{\alpha_3}
\sum_{\beta=1}^d
\frac{\partial X_{\alpha_1}}{\partial x_{\beta}}
\frac{\partial X_{\alpha_2}}{\partial x_{\beta}}$,
we may express the element tensor $A^K$ as the tensor contraction
\begin{equation}
  A^K_i = \sum_{\alpha} A^0_{i\alpha} G_K^{\alpha}.
\label{eq:tensorrepresentation}
\end{equation}
We refer to $A^0$ as the \emph{reference tensor} and to $G_K$ as the
\emph{geometry tensor}.

The main point of this representation is that the reference
tensor~$A^0$ is independent of the triangulation~$\mathcal{T}$
and may thus be precomputed. During assembly, one may then
iterate over all elements of the triangulation and on each
element~$K$ compute the geometry tensor~$G_K$, compute the
tensor contraction~(\ref{eq:tensorrepresentation}) and then
add the resulting element tensor~$A^K$ to the global sparse
matrix~$A$. The form compiler FFC automatically generates the tensor
representation~(\ref{eq:tensorrepresentation}); i.e., it precomputes the
reference tensor~$A^0$ at compile-time and generates code for evaluating
the geometry tensor~$G_K$ and the tensor contraction.
\section{Extending the framework to discontinuous Galerkin methods}
\label{sec:dg}

\begin{figure}
  \begin{center}
    \psfrag{K1}{$K^{+}$}
    \psfrag{K2}{$K^{-}$}
    \psfrag{S}{$S$}
    \includegraphics[width=4cm]{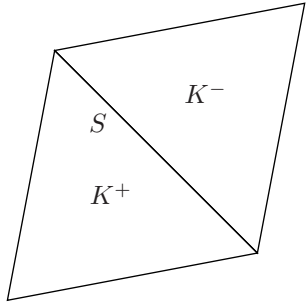}
  \end{center}
\caption{Two cells $K^{+}$ and $K^{-}$ sharing a common facet $S$.}
\label{fig:twocells}
\end{figure}

To extend the above framework for finite element assembly to discontinuous
Galerkin methods, we need to consider variational forms expressed
as integrals over the \emph{interior facets} of a finite element
mesh. Consider for example the following bilinear form which may appear
as a term in a discontinuous Galerkin formulation:
\begin{equation}
  a(v, u)
   = \sum_{S \in \partial_i \mathcal{T}} \int_S \jump{v} \jump{u} \ds,
 \label{eq:dgexample}
\end{equation}
where $\partial_i \mathcal{T}$ denotes the set of all interior facets
of the triangulation~$\mathcal{T}$ and where $\jump{v}$ denotes the
\emph{jump} in the function value of~$v$ across the facet~$S$:
\begin{equation}
  \jump{v} = v^{+} - v^{-}.
\end{equation}
Here, $v^{+}$ and $v^{-}$ denote the values of $v$ on the facet $S$
as seen from the two cells $K^{+}$ and $K^{-}$ incident with $S$,
respectively (see Figure~\ref{fig:twocells}). We note that each interior
facet is incident to exactly two cells, and we may label these two cells
$K^{+}$ and~$K^{-}$.
%
\subsection{A general assembly algorithm}
To assemble the global sparse tensor~$A$ for variational forms that
contain integrals over interior facets as in~(\ref{eq:dgexample}),
one may extend the standard assembly algorithm over the cells of the
computational mesh by including an iteration over the interior facets
of the mesh. Similarly, one may extend the assembly to include the
\emph{exterior facets} (the set of facets incident with $\partial \Omega$)
to account for terms that involve integrals over the boundary of the mesh.

A general assembly algorithm for the computation of the global sparse
tensor $A$ is given in~\cite{logg:manual:04}. This algorithm iterates
first over all the cells of the mesh to compute the local contribution
from each cell to the global sparse tensor. Above, we referred to
this contribution as the element tensor~$A^{K}$. In the context of the
general assembly algorithm we shall refer to the contribution from each
cell as the \emph{cell tensor}~$A^{K}$.  Similarly, one may iterate
over the exterior and interior facets of the mesh and add the local
contribution from each facet to the global sparse tensor. We refer to
these local contributions, denoted by $A^{S}$, as the \emph{exterior}
or \emph{interior} \emph{facet tensors}.
\subsection{Computing the interior facet tensor}
To define the interior facet tensor~$A^{S}$ for a given multilinear
form~$a$ expressed as an integral over the set of interior
facets~$\partial_i\Omega$ such as in~(\ref{eq:dgexample}), we write
\begin{equation}
  A_i = a(\phi_{i_1}, \phi_{i_2})
  = \sum_S
  a_S(\phi_{i_1}, \phi_{i_2}),
\label{eq:assembly,S}
\end{equation}
where the summation is carried out only over those interior facets
where both $\phi_{i_1}$ and $\phi_{i_2}$ are nonzero. In the case
of~(\ref{eq:dgexample}), we have $a_S(v, u) = \int_S \jump{v} \jump{u}
\ds$. To assemble the global sparse tensor~$A$ efficiently, one may
introduce a \emph{local-to-global} mapping that maps the basis functions
on the local facet $S$ to the set of global basis functions. To
construct this mapping, consider again two cells $K^+$ and $K^-$
sharing a common facet $S$ as in Figure~\ref{fig:twocells}. As above,
let $\{\phi_i\}_{i=1}^{N}$ be a global (possibly discontinuous) basis
for $V_h$. For ease of notation, we consider the assembly of the global
tensor (matrix) for a bilinear form for $V_h^1 = V_h^2 = V_h$ and drop
the index $j$ (see equation~(\ref{eq:generic_multilinear_form})). We
thus assume here that all discretising function spaces are equal, but
note that this is not necessary. In particular, our implementation in
\ffc{} does not make this assumption and is able to generate code for
assembly of tensors of arbitrary ranks for arbitrary combinations of
finite element function spaces.

Furthermore, let $\{\phi^{K^+}_i\}_{i=1}^{n}$ be the local finite element
basis on $K^+$ and let similarly $\{\phi^{K^-}_i\}_{i=1}^{n}$ be the local
finite element basis on $K^-$. We now extend these local basis functions
to the macro cell $\bar{K} = K^+ \cup K^-$ by the following construction:
\begin{equation}
  \bar{\phi}^{\bar{K}}_i(x) =
  \left\{
  \begin{array}{lll}
    \phi_i^{K^+}(x),     \quad & i = 1,2,\ldots, n, \quad & x \in K^+, \\
    0,                   \quad & i = 1,2,\ldots, n, \quad & x \in K^-, \\
    0,                   \quad & i = n+1,n+2,\ldots, 2n, \quad & x \in K^+, \\
    \phi_{i-n}^{K^-}(x), \quad & i = n+1,n+2,\ldots, 2n, \quad & x \in K^-. \\
  \end{array}
  \right.
\label{eq:macrobasis}
\end{equation}
We thus extend the local basis functions on $K^+$ and $K^-$ to $\bar{K}$
by zero to obtain a local finite element space on $\bar{K}$ of dimension
$2n$.  For each $K \in \mathcal{T}$, we further let
\begin{equation}
  \iota_K : [1,n] \rightarrow [1,N]
\label{eq:iota_K}
\end{equation}
be a standard \emph{local-to-global} mapping, that is, a mapping from a
local enumeration of the basis functions on cell~$K$ to the corresponding
global basis functions such that $\phi^K_i = \phi_{\iota_K(i)}\vert_K$
for $i = 1,2,\ldots,n$.  By the construction~(\ref{eq:macrobasis}),
we obtain a local-to-global mapping for $\bar{K}$ (or $S$). Thus,
$\iota_{\bar{K}}(1) = \iota_{{K^+}}(1),
\ldots,
\iota_{\bar{K}}(n) = \iota_{{K^+}}(n),
\iota_{\bar{K}}(n+1) = \iota_{{K^-}}(1),
\ldots, \iota_{\bar{K}}(2n) = \iota_{{K^-}}(n)$.

We may now proceed to define the interior facet tensor $A^S$.  Consider
first the case when $\iota_{\bar{K}}$ is an injective mapping and note
that $\iota_{\bar{K}}$ is injective when the ranges of $\iota_{K^+}$
and $\iota_{K^-}$ are disjoint (which is the case for discontinuous
elements). Continuing from~(\ref{eq:assembly,S}), we then obtain
\begin{equation}
  A_i
  = \sum_S
  a_S(\phi_{i_1}, \phi_{i_2})
  = \sum_S
  a_S
  \left(
  \bar{\phi}^{\bar{K}}_{\iota_{\bar{K}}^{-1}(i_1)},
  \bar{\phi}^{\bar{K}}_{\iota_{\bar{K}}^{-1}(i_2)}
  \right)
  = \sum_S
  A^S_{\iota_{\bar{K}}^{-1}(i_1), \iota_{\bar{K}}^{-1}(i_2)},
\end{equation}
where the interior facet tensor~$A^S$ is thus defined by
\begin{equation}
  A^S_i = a_S(\bar{\phi}^{\bar{K}}_{i_1}, \bar{\phi}^{\bar{K}}_{i_2}), \quad
  i_1, i_2 = 1, 2, \ldots, 2n.
\label{eq:interiorfacettensor}
\end{equation}
It follows that the global tensor~$A$ may be computed by
Algorithm~\ref{alg:interiorassembly}.
\begin{algorithm}
  \begin{center}
    \begin{tabbing}
      $A = 0$ \\
      \textbf{for} $S \in \partial_i \mathcal{T}$ \\
      \tab \textbf{for} $i_1, i_2 = 1, 2, \ldots, 2n$ \\
      \tab \tab $A^S_i = a_S(\bar{\phi}^{\bar{K}}_{i_1},
      \bar{\phi}^{\bar{K}}_{i_2})$ \\
      \tab \textbf{end for} \\
      \tab \textbf{for} $i_1, i_2 = 1, 2, \ldots, 2n$ \\
      \tab \tab $A_{\iota_{\bar{K}}(i_1), \iota_{\bar{K}}(i_2)} +\!\!= A^S_{i_1, i_2}$ \\
      \tab \textbf{end for} \\
      \textbf{end for}
    \end{tabbing}
  \end{center}
\caption{Assembly algorithm over interior facets.}
\label{alg:interiorassembly}
\end{algorithm}

Now, if $\iota_{\bar{K}}$ is not injective (two local basis functions
are restrictions of the same global basis function), which may happen
if the basis functions are continuous, we may still assemble the global
tensor~$A$ by Algorithm~\ref{alg:interiorassembly} and compute the
interior facet tensor as in~(\ref{eq:interiorfacettensor}).  To see
this, assume that $\iota_{\bar{K}}(i_1) = \iota_{\bar{K}}(i_1') =
I$ for some $i_1 \neq i_1'$. It then follows that the entry $A_{I,
\iota_{\bar{K}}(i_2)}$ will be a sum of the two terms $A^S_{i_1,i_2}$
and $A^S_{i_1',i_2}$ (and possibly other terms). Since $a_S$ is bilinear,
we have
\begin{equation}
  A^S_{i_1,i_2} + A^S_{i_1',i_2}
  =
  a_S(\bar{\phi}^{\bar{K}}_{i_1}, \bar{\phi}^{\bar{K}}_{i_2}) +
  a_S(\bar{\phi}^{\bar{K}}_{i_1'}, \bar{\phi}^{\bar{K}}_{i_2})
  =
  a_S(\bar{\phi}^{\bar{K}}_{i_1} + \bar{\phi}^{\bar{K}}_{i_1'},
  \bar{\phi}^{\bar{K}}_{i_2})
  =
  a_S(\phi_I, \bar{\phi}^{\bar{K}}_{i_2}),
\end{equation}
where by the construction~(\ref{eq:macrobasis}) $\phi_I$ is the
global basis function that both $\bar{\phi}^{\bar{K}}_{i_1}$ and
$\bar{\phi}^{\bar{K}}_{i_1'}$ are mapped to.
\subsection{Tensor representation and precomputation on facets}
\label{sec:tensorrepresentation}

In Section~\ref{sec:compiling_forms}, we described how the cell tensor
(element tensor) may be computed from the tensor representation
\begin{equation} \label{eq:G1}
  A^K_i = \sum_{\alpha} A^0_{i\alpha} G_K^{\alpha}.
\end{equation}
Similarly, one may use the affine mappings $F_{K^+}$ and
$F_{K^-}$ to obtain a tensor representation for the interior facet
tensor~$A^S$. However, depending on the topology of the macro cell
$\bar{K}$, one obtains different tensor representations. For a triangular
mesh, each cell has three facets (edges) and there are thus $3 \times 3
= 9$ different topologies to consider; there are nine different ways in
which two edges can meet. Similarly, for a tetrahedral mesh, there are
$4 \times 4 = 16$ different topologies to consider.\footnote{FFC assumes
a particular local ordering of the entities of the computational mesh
as described in~\cite{logg:manual:02}. If no particular ordering of the
mesh entities is assumed, one needs to consider $3 \times 3 \times 2 =
18$ different topologies for triangles and $4 \times 4 \times 6 = 96$
topologies for tetrahedra. This is because there are two different ways to
superimpose two edges, and there are six ways to superimpose two faces.}
We thus obtain a tensor representation of the form
\begin{equation} \label{eq:G2}
  A^S_i = \sum_{\alpha} A^{0,f^+(S),f^-(S)}_{i\alpha} G_{\bar{K}(S)}^{\alpha},
\end{equation}
where $f^+$ and $f^-$ denote the local numbers of the two facets that
meet at $S$ relative to the two cells $K^+$ and $K^-$ respectively.
We note that the geometry tensor $G_K^{\alpha}$ in~(\ref{eq:G1})
involves the mapping from the reference cell and differs from the geometry
tensor $G_{\bar{K}(S)}^{\alpha}$ in~(\ref{eq:G2}), which may involve the
mapping from the reference cell and the mapping from the reference facet.
The form compiler FFC precomputes the reference tensor $A^{0,f^+,f^-}$
for each facet--facet combination $(f^+, f^-)$ and a run-time decision
must be made as to which reference tensor should be contracted with the
geometry tensor.
\section{Implementation}
\label{sec:implementation}

We give here a short introduction to the \ffc{} language before putting
the compiler into context with respect to the other components of
\fenics{}. Then we discuss the performance in terms of the efficiency of
the generated code as well as the benefits of automated code generation
in general.
\subsection{The Form Compiler FFC}

The form compiler FFC computes the tensor
representation~(\ref{eq:tensorrepresentation}) from a high-level
description of the variational form, and generates efficient C++ code for
the computation of the element tensor based on this representation. It
is also possible to generate code that uses the conventional quadrature
representation. Code can be generated which is consistent with the
UFC~\cite{logg:manual:04} specification, although any format in any
language can be implemented.

The bilinear form for the weighted Laplacian~(\ref{eq:weightedlaplacian})
can be expressed in the FFC form language by \texttt{a = w*dot(grad(v),
grad(u))*dx}. Integration over a cell is denoted by \texttt{*dx},
integration over an exterior facet is denoted by \texttt{*ds} and
integration over an interior facet is denoted by~\texttt{*dS}.  The \ffc{}
form language is equipped with basic differential operators including
partial derivatives, \texttt{v.dx(i)}; the gradient, \texttt{grad(v)};
the divergence, \texttt{div(v)}; and the curl, \texttt{curl(v)}. Basic
linear algebra operators like inner products \texttt{dot(v, w)} and
matrix-vector multiplications \texttt{mult(A, v)} are also implemented.
Functions which are evaluated on facets can be `restricted' to the plus
or minus sides of the facet. A function $v$ evaluated on the plus side of
a facet is expressed as \texttt{v('+')}, and the same function evaluated
on the minus side is expressed as~\texttt{v('-')}. Typical discontinuous
Galerkin operators are available such as \texttt{jump(v)}, which is
equivalent to $v^{+} - v^{-}$; \texttt{jump(v, n)}, which is equivalent
to $v^{+} \vect{n}^{+} + v^{-} \vect{n}^{-}$ or $\vect{v}^{+} \cdot
\vect{n}^{+} + \vect{v}^{-} \cdot \vect{n}^{-}$; and \texttt{avg(v)},
which is equivalent to $\brac{v^{+} + v^{-}}/2$.

\ffc{} is written in Python, and the interface to \ffc{} also uses
the Python syntax which makes the addition of user-defined operators
simple. This will be demonstrated in Section~\ref{sec:examples}, as will
be the use of the operators introduced above.
\subsection{FEniCS}

The form compiler \ffc{} is a core component of FEniCS~\cite{fenics},
a software system aiming at automation of various aspects of
computational mathematical modelling, in particular the solution
of partial differential equations. Other core components of FEniCS
include FIAT~\cite{Kir04,Kir05}, SyFi~\cite{www:SyFi,Mardal.2006.6},
UFC~\cite{AlnLog2008} and DOLFIN~\cite{www:dolfin}.

FIAT is a tool for generating and tabulating finite element basis
functions for a range of finite element spaces. FFC calls FIAT at
compile-time to evaluate basis functions on the reference element as
described in Section~\ref{sec:tensorrepresentation}.  FFC supports the
generation of C++ code which is consistent with the Unified Form-assembly
Code (UFC) specification.  Any library which supports the UFC interface
can use the automatically generated code to assemble forms.

Finally, DOLFIN is a consistent high-level problem-solving environment
(PSE) for the solution of partial differential equations.  DOLFIN handles
the communication between the core components of FEniCS. Amongst other
things, DOLFIN manages meshes (it does not generate meshes) and provides
various linear algebra solvers and tools, as well as an interface to
specialised linear algebra libraries. DOLFIN supports the UFC interface
and uses the code generated by FFC to assemble and solve the global
linear systems.
\subsection{Performance}
\label{sec:performance}

As claimed above, the form compiler FFC generates efficient code that may
compete with and in some cases outperforms hand-written and optimised
code. This was demonstrated in~\cite{logg:article:10} where speedups
as large as a factor~1000 were demonstrated for a set of standard test
cases.\footnote{It should be noted that these
  speedups concern the evaluation of the element tensor on each local
  element. Insertion into the global sparse matrix and solution of the
  linear system are not included. The global speedup is smaller and
  depends on the efficiency of linear algebra and mesh data structures,
  as well as the problem at hand, the efficiency of the linear solver,
  and choice of preconditioner.}
These speedups are a consequence of the reduced complexity of computing
the tensor contraction~(\ref{eq:tensorrepresentation}) compared to
a run-time iteration over quadrature points for certain forms. It was
further demonstrated in~\cite{logg:article:07,logg:article:09,KirLog2008a}
that these results may be further improved by finding \emph{a priori}
so-called \emph{complexity-reducing relations} between subtensors of
the reference tensor~$A^0$.

As an example, consider here the operation count for evaluating the
tensor-contraction for the Laplacian,
\begin{equation}
  A^K_i = \sum_{\alpha_1=1}^d \sum_{\alpha_2=1}^d
  \det F_K'
  \sum_{\beta=1}^d
  \frac{\partial X_{\alpha_1}}{\partial x_{\beta}}
  \frac{\partial X_{\alpha_2}}{\partial x_{\beta}}
  \int_{K_0}
  \frac{\partial \Phi_{i_1}}{\partial X_{\alpha_1}}
  \frac{\partial \Phi_{i_2}}{\partial X_{\alpha_2}}
  \dX.
\end{equation}
With $n$ the dimension of the local finite element function space, each
of the $n^2$ entries of the element tensor~$A^K$ may be evaluated in
roughly $T_T \sim d^2(d + 1) + d^2 \sim d^3$ operations; first $d + 1$
operations to evaluate each entry of the $d \times d$ geometry tensor
$G_K$ and then $d^2$ operations to perform the tensor contraction.

If instead we use a loop over $N$ quadrature points at run-time,
each entry of $A^K$ may be evaluated in $CN$ operations, where $C$
is the cost of evaluating the integrand at each quadrature point. It
is difficult to estimate exactly the size of $C$, but a reasonable
estimate is $2d^2$ (transforming the gradient of both the test and trial
functions from the reference element by a matrix-vector product with
the inverse transpose of the Jacobian matrix). It is assumed that the
values of the gradients of all basis functions have been pretabulated
at the quadrature points on the reference element.  Now, the number of
quadrature points needed for exact integration is $N \sim (2(k - 1) +
1)^d = (2k-1)^d$, where $k$ is the polynomial degree of the finite element
basis functions. It follows that the complexity of quadrature is $T_Q
\sim CN \sim 2d^2 (2k-1)^d$.  We thus find that the speedup of the tensor
contraction~(\ref{eq:tensorrepresentation}) compared to quadrature is
\begin{equation}
  \frac{T_Q}{T_T} \sim \frac{2d^2 (2k-1)^d}{d^3}
  = 2(2k-1)^d/d.
\end{equation}
It follows that the speedup may be substantial for moderate sized values
of $k$. For $k = 1$, it is not clear that the speedup is positive,
although it was demonstrated in~\cite{logg:article:10} that the speedup
in this case is a factor~$\sim 10$.

If instead we consider the weighted
Laplacian~(\ref{eq:weightedlaplacian}), it is less clear
that the tensor contraction outperforms quadrature. As discussed
in~\cite{logg:article:10}, the relative performance of quadrature improves
with an increasing number of coefficient functions (weights) in the
variational form. Because of this, the form compiler FFC supports a mode
where quadrature code is generated instead of code based on the tensor
representation~(\ref{eq:tensorrepresentation}).  Even with quadrature,
various simple \emph{a prior} optimisations are possible when using
precomputation, such as the elimination of operations on zero entries
and loop unrolling.

In summary, the advantage of code generation is not that the tensor
representation~(\ref{eq:tensorrepresentation}) always leads to more
efficient code than hand-written code, in particular
since~(\ref{eq:tensorrepresentation}) may (with some effort) also be
coded manually. Instead, the merit lies in:~(i) the potential to
employ sophisticated strategies for evaluation of the element tensor
as in~\cite{logg:article:10}; (ii) the potential to employ
sophisticated compile-time optimisation techniques as
in~\cite{logg:article:07,logg:article:09,KirLog2008a}; (iii) the
generation of architecture-specific code (in particular for multicore
processors); and (iv) the reduction in development time through
simple and compact coding of finite element variational forms
while retaining efficiency. In particular, the simplicity with
which forms can be coded is demonstrated in the following section.
For all examples in the following section, the time required for
FFC to generate C++ code is of the order of seconds. The quantity of
automatically produced code is highly dependent on the complexity
of the considered form.
\section{Examples}
\label{sec:examples}

We demonstrate the compilation of variational forms for discontinuous
Galerkin methods through a number of examples.

\subsection{Poisson's equation}
Consider the function space $V_{h}$,
\begin{equation}
  V_{h} = \bracc{v \in L^{2}\brac{\Omega}: \ v\vert_K \in P_{k}\brac{K}
   \forall K \in \mathcal{T}},
\label{eq:space_scalar_L2}
\end{equation}
where $P_{k}\brac{K}$ denotes the space of polynomials of degree $k$ on
the element $K$. Setting $V^{1}_{h} = V^{2}_{h} = V_{h}$, the bilinear
and linear forms for the Poisson equation with homogeneous Dirichlet
boundary conditions read~\cite{arnold_et_al:2002}
\begin{multline}
  a(v,u) =
    \int_\Omega \nabla v \cdot \nabla u \; \dx
  - \int_{\Gamma^{0}} \jump{v} \cdot \avg{\nabla u} \; \ds
  - \int_{\Gamma^{0}} \avg{\nabla v} \cdot \jump{u} \; \ds
\\
  - \int_{\partial\Omega} \jump{v} \cdot \nabla u \; \ds
  - \int_{\partial\Omega} \nabla v \cdot \jump{u} \; \ds
  + \int_{\Gamma^{0}} \frac{\alpha}{h} \jump{v} \cdot \jump{u} \; \ds
  + \int_{\partial\Omega} \frac{\alpha}{h} v u \; \ds
\label{eq:poisson_bilinear}
\end{multline}
and
\begin{equation}
  L(v) = \int_{\Omega} v f \; \dx,
\label{eq:poisson_linear}
\end{equation}
where $\Gamma^0$ denotes interior facets, $\alpha$ is a penalty parameter
and $h$ is a measure for the average of the mesh size defined as $h =
(h^{+} + h^{-})/2$ for the two cells $K^{+}$ and $K^{-}$ incident with
the given interior facet. The size of a cell is defined here as twice the
circumradius.  The jump $\jump{\cdot}$ and average $\avg{\cdot}$ operators
are defined as $\jump{v} = v^{+} \vect{n}^{+} + v^{-} \vect{n}^{-}$
and $\avg{\nabla v} = (\nabla v^{+} + \nabla v^{-})/2$ on $\Gamma^{0}$
and $\jump{v} = v \vect{n}$ on $\partial\Omega$. Here, $\vect{n}^{+}$ and
$\vect{n}^{-}$ denote the outward unit normal to the given facet as seen
from the two cells $K^{+}$ and $K^{-}$ respectively.  The corresponding
\ffc{} input for this problem is shown in Table~\ref{tab:poisson_equation}
for 5th order polynomials on triangular elements. The form and syntax
of the compiler input resembles closely the mathematical notation
in~(\ref{eq:poisson_bilinear}) and~(\ref{eq:poisson_linear}).  Note that
in the code we need to restrict the (potentially) multi-valued function
$h$ to either $K^+$ or $K^-$ (here \texttt{h('+')}) even if the function
in this particular case is single-valued ($h = (h^+ + h^-)/2$).
\begin{table}
\caption{\ffc{} input for the interior penalty method applied to the
         Poisson equation using~$k=5$.}
\label{tab:poisson_equation}
\begin{code}{0.8}
element = FiniteElement("Discontinuous Lagrange", "triangle", 5)

v = TestFunction(element)
u = TrialFunction(element)
f = Function(element)

n = FacetNormal("triangle")
h = MeshSize("triangle")

alpha = 32.0

a =  dot(grad(v), grad(u))*dx \
   - dot(jump(v, n), avg(grad(u)))*dS \
   - dot(avg(grad(v)), jump(u, n))*dS \
   - dot(mult(v, n), grad(u))*ds \
   - dot(grad(v), mult(u, n))*ds \
   + alpha/h('+')*dot(jump(v, n), jump(u, n))*dS \
   + alpha/h*v*u*ds

L = v*f*dx
\end{code}
\end{table}
\subsection{Advection--diffusion equation}

We consider now the advection--diffusion equation with homogeneous
Dirichlet boundary conditions on inflow boundaries and full upwinding
of the advective flux at element facets. Setting $V^{1}_{h} = V^{2}_{h}
= V_{h}$, where $V_{h}$ is defined as in~(\ref{eq:space_scalar_L2}),
the bilinear and linear forms read
\begin{multline}
  a(v,u) =
     \int_{\Omega} \nabla v \cdot \brac{ \kappa \nabla u - \vect{b} u} \; \dx
    + \int_{\Gamma^{0}} \jump{v} \cdot \vect{b} u^{\star} \; \ds
    + \int_{\partial\Omega} \jump{v} \cdot \vect{b} u^{\star} \; \ds
\\
    - \int_{\Gamma^{0}} \kappa \jump{v} \avg{\nabla u} \; \ds
    - \int_{\Gamma^{0}} \kappa \avg{\nabla v} \cdot \jump{u} \; \ds
    - \int_{\partial\Omega} \kappa \jump{v} \cdot \nabla u \; \ds
\\
    - \int_{\partial\Omega} \kappa \nabla v \cdot \jump{u} \; \ds
    + \int_{\Gamma^{0}} \frac{\kappa\alpha}{h} \jump{v} \cdot \jump{u} \; \ds
    + \int_{\partial\Omega} \frac{\kappa\alpha}{h} v u \; \ds
\label{eq:advection-diffusion_bilinear}
\end{multline}
and
\begin{equation}
\label{eq:advection-diffusion_linear}
  L(v) = \int_{\Omega} v f \; \dx,
\end{equation}
where the vector $\vect{b}$ is a given velocity field, $u^{\star}$
is equal to $u$ restricted to the upwind side of a facet,
\begin{equation}
  u^{\star} =
  \begin{cases}
    u^{+}      & \vect{b} \cdot \vect{n^{+}} \geq 0, \\
    u^{-}      & \vect{b} \cdot \vect{n^{+}} < 0,
  \end{cases}
\label{eq:upwind_value}
\end{equation}
and $\kappa$ is the diffusion coefficient. The definitions of the jump and
average operators are the same as for the Poisson equation.  The \ffc{}
input for this problem is depicted in Table~\ref{tab:advection_diffusion},
and is again a reflection of the mathematical formulation.
\begin{table}
\caption{\ffc{} input for the advection--diffusion equation using $k=3$.}
\label{tab:advection_diffusion}
\begin{code}{0.9}
scalar   = FiniteElement("Discontinuous Lagrange", "triangle", 3)
vector   = VectorElement("Lagrange", "triangle", 3)
constant = FiniteElement("Discontinuous Lagrange", "triangle", 0)

v  = TestFunction(scalar)
u  = TrialFunction(scalar)

b  = Function(vector)
f  = Function(scalar)
n  = FacetNormal("triangle")
h  = MeshSize("triangle")
of = Function(constant)

kappa = 0.2
alpha = 20.0

def upwind(b, u):
  return [b[i]('+')*(of('+')*u('+') + of('-')*u('-')) for i in range(len(b))]

a =  dot(grad(v), mult(kappa, grad(u)) - mult(b, u))*dx \
   + dot(jump(v, n), upwind(b, u))*dS \
   + dot(mult(v, n), mult(b, of*u))*ds \
   - kappa*dot(jump(v, n), avg(grad(u)) )*dS \
   - kappa*dot(avg(grad(v)), jump(u, n))*dS \
   - kappa*dot(mult(v, n), grad(u))*ds \
   - kappa*dot(grad(v), mult(u, n))*ds \
   + kappa*alpha/h('+')*dot(jump(v, n), jump(u, n))*dS \
   + kappa*alpha/h*v*u*ds

L = v*f*dx
\end{code}
\end{table}
In this particular implementation, the same basis has been used for
the solution and components of the advective velocity field, although
different orders can be used.  The value of the discontinuous function
`\texttt{of}' (outflow facet) in Table~\ref{tab:advection_diffusion}
is either 1 or 0 on a side of the interior facet which is being
considered. When performing a computation, we compute this value in
\dolfin{} according to the definition in~(\ref{eq:upwind_value}) and
pass it to the form as a function.

If a facet is an outflow facet relative to the element~$K^{+}$, i.e.,
$\vect{b} \cdot \vect{n}^{+} \geq 0$, then the value of \texttt{of('+')}
is 1, while the value of \texttt{of('-')} is 0 and vice versa.  As a
consequence, the return value of the `\texttt{upwind}' function is equal
to $\vect{b} u^{\star}$. The `\texttt{upwind}' function is an example
of how one can extend the \ffc{} language with user-defined functions
written in Python.
\subsection{The Stokes equations}
We consider next the Stokes equations with a mixture of continuous
and discontinuous functions, as well as basis functions with possibly
varying polynomial orders. Consider the function spaces~$\vect{V}_{h}$
and~$Q_{h}$,
\begin{align}
  \vect{V}_{h} &= \bracc{\vect{v} \in \brac{L^{2}\brac{\Omega}}^{d}:
                  \ v_{i} \in P_{k}\brac{K}  \forall K \in \mathcal{T}, \ 1 \leq i \leq d},\\
  Q_{h} &= \bracc{q \in H^{1} \brac{\Omega}: \ q \in P_{j}\brac{K}
                  \forall K \in \mathcal{T}}.
\end{align}
Setting $V^{1}_{h} = V^{2}_{h} = \vect{V}_{h} \times Q_{h}$ and $\vect{u}
= 0$ on $\partial \Omega$, particular bilinear and linear forms for the
Stokes equation read~\cite{baker:1990}
\begin{multline}
  a(\vect{v},q;\vect{u}, p) =
         \int_\Omega \nu \nabla \vect{v}  : \nabla \vect{u} \; \dx
         + \int_{\Omega} \vect{v} \cdot \nabla p \; \dx
         - \int_{\Omega} \nabla q \cdot \vect{u} \; \dx
\\
         + \int_{\Gamma^{0}} q \jump{\vect{u} \cdot \vect{n}} \; \ds
         + \int_{\partial \Omega} q \vect{u} \cdot \vect{n} \; \ds
\\
         - \int_{\Gamma^{0}} \nu \jump{\vect{v}} : \avg{\nabla \vect{u}} \; \ds
         - \int_{\Gamma^{0}} \nu \avg{\nabla \vect{v}} : \jump{\vect{u}} \; \ds
         - \int_{\partial \Omega} \nu \jump{\vect{v}} : \nabla \vect{u} \; \ds
\\
         - \int_{\partial \Omega} \nu \nabla \vect{v} : \jump{\vect{u}} \; \ds
         + \int_{\Gamma^{0}} \frac{\nu \alpha}{h} \jump{\vect{v}} : \jump{\vect{u}} \; \ds
         + \int_{\partial \Omega} \frac{\nu \alpha}{h} \jump{\vect{v}} : \jump{\vect{u}} \; \ds,
\label{eq:stokes_bilinear}
\end{multline}
and
\begin{equation}
  L(\vect{v}, q) = \int_\Omega \vect{v} \cdot \vect{f} \; \dx.
\label{eq:stokes_linear}
\end{equation}
The jump $\jump{\cdot}$ and average $\avg{\cdot}$ operators are
defined as $\jump{\vect{v}} = \vect{v}^{+} \otimes \vect{n}^{+} +
\vect{v}^{-} \otimes \vect{n}^{-}$, $\jump{\vect{v} \cdot \vect{n}} =
\vect{v}^{+} \cdot \vect{n}^{+} + \vect{v}^{-} \cdot \vect{n}^{-}$ and
$\avg{\nabla \vect{v}} = (\nabla \vect{v}^{+} + \nabla \vect{v}^{-})/2$
on $\Gamma^{0}$ and $\jump{\vect{v}} = \vect{v} \otimes \vect{n}$
on $\partial\Omega$.  The \ffc{} input for this problem with $k=j=1$,
as proposed in~\cite{baker:1990}, and the kinematic viscosity $\nu=1.0$
is shown in Table~\ref{tab:stokes}.
\begin{table}
\caption{\ffc{} input for the Stokes equation using~$k=1$.}
\label{tab:stokes}
\begin{code}{0.9}
V = VectorElement("Discontinuous Lagrange", "triangle", 1)
Q = FiniteElement("Lagrange", "triangle", 1)

element = V + Q

(v, q) = TestFunctions(element)
(u, p) = TrialFunctions(element)

f = Function(V)
n = FacetNormal("triangle")
h = MeshSize("triangle")

alpha = 4.0

a =  dot(grad(v), grad(u))*dx + dot(v, grad(p))*dx - dot(grad(q), u)*dx \
   + dot(q('+'), jump(u, n))*dS \
   + q*dot(u, n)*ds \
   - dot(mult(avg(grad(v)), n('+')), jump(u))*dS \
   - dot(jump(v), mult(avg(grad(u)), n('+')))*dS \
   - dot(mult(grad(v), n), u)*ds \
   - dot(v, mult(grad(u), n))*ds \
   + alpha/h('+')*dot(jump(v), jump(u))*dS \
   + alpha/h*dot(v, u)*ds

L = dot(v, f)*dx
\end{code}
\end{table}
\subsection{Biharmonic equation}

Classically, Galerkin methods for the biharmonic equation seek
approximate solutions in a subspace of $H^{2}\brac{\Omega}$. However,
such functions are difficult to construct in a finite element
context. Based on discontinuous Galerkin principles, methods have
been developed which utilise functions from $H^{1}\brac{\Omega}$
\cite{engel:2002,wells:2007}. Rather than considering jumps in functions
across element boundaries, terms involving the jump in the normal
derivative across element boundaries are introduced.  Unlike fully
discontinuous approaches, this method does not involve double-degrees of
freedom on element edges and therefore does not lead to the significant
increase in the number of degrees of freedom relative to conventional
methods. Consider the continuous function space
\begin{equation}
  V_{h} = \bracc{v \in H^{1}_{0}\brac{\Omega}: \ v \in P_{k}\brac{K}
    \forall K \in \mathcal{T}}.
\end{equation}
Setting $V^{1}_{h} = V^{2}_{h} = V_{h}$, the bilinear and linear forms for
the biharmonic equation, with the boundary conditions $u = 0 \ {\rm on} \
\partial \Omega$ and $\nabla^{2} u = 0 \ {\rm on} \ \partial \Omega$, read
\begin{multline}
  a(v,u) =
      \int_\Omega \nabla^{2} v  \nabla^{2} u \; \dx
    - \int_{\Gamma^{0}} \jump{\nabla v} \cdot \avg{\nabla^{2} u} \; \ds
    - \int_{\Gamma^{0}} \avg{\nabla^{2} v} \cdot \jump{\nabla u} \; \ds \\
    + \int_{\Gamma^{0}} \frac{\alpha}{h} \jump{\nabla v} \cdot \jump{\nabla u} \; \ds,
\label{eq:biharmonic_bilinear}
\end{multline}
\begin{equation}
  L(v) = \int_\Omega v f \; \dx.
\label{eq:biharmonic_linear}
\end{equation}
The jump $\jump{\cdot}$ and average $\avg{\cdot}$ operators are defined as
$\jump{\nabla v} = \nabla v^{+} \cdot \vect{n}^{+} + \nabla v^{-} \cdot
\vect{n}^{-}$ and $\avg{\nabla^{2} v} = (\nabla^{2} v^{+} + \nabla^{2}
v^{-})/2$ on $\Gamma^{0}$.  The \ffc{} input for this problem with $k=4$
is shown in Table~\ref{tab:biharmonic}.
\begin{table}
\caption{\ffc{} input for the biharmonic equation using $k=4$.}
\label{tab:biharmonic}
\begin{code}{0.8}
element = FiniteElement("Lagrange", "tetrahedron", 4)

v = TestFunction(element)
u = TrialFunction(element)
f = Function(element)

n = FacetNormal("tetrahedron")
h = MeshSize("tetrahedron")

alpha = 16.0

a =  dot(div(grad(v)), div(grad(u)))*dx \
   - dot(jump(grad(v), n), avg(div(grad(u))))*dS \
   - dot(avg(div(grad(v))), jump(grad(u), n))*dS \
   + alpha/h('+')*dot(jump(grad(v), n), jump(grad(u), n))*dS

L = v*f*dx
\end{code}
\end{table}

\begin{figure}
  \begin{center}
    \input{convergence_biharmonic.tex}
  \end{center}
\caption{Error in the $L^{2}$ norm for the biharmonic equation with penalty parameters
         $\alpha=4$, $\alpha=16$ and $\alpha=16$,
         for $k=2$, $k=3$ and $k=4$ respectively.}
\label{fig:biharmonic_convergence}
\end{figure}
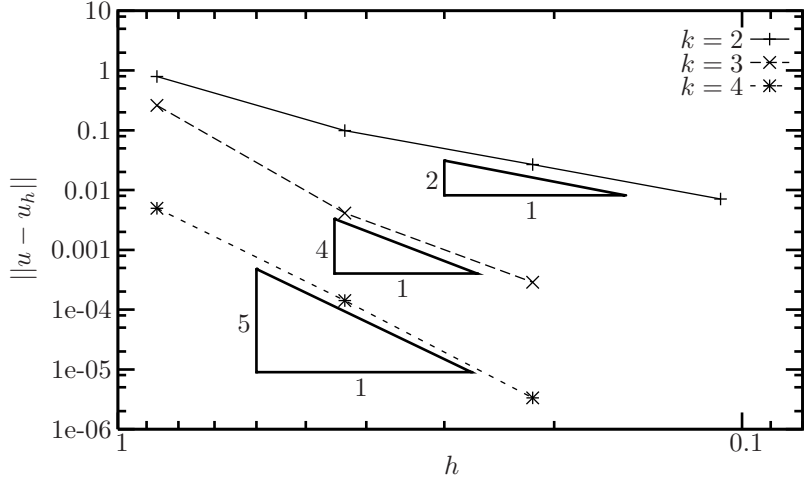

For the biharmonic equation we consider an example on the domain
$\Omega = [0,1] \times [0,1] \times [0,1]$ with $f = 9 \pi^{4}
\sin(\pi x) \sin(\pi y) \sin(\pi z)$, in which case the exact solution
$u = \sin(\pi x) \sin(\pi y) \sin(\pi z)$. The observed convergence
behaviour is illustrated in Figure~\ref{fig:biharmonic_convergence} for
various polynomial orders. As predicted by \emph{a priori} estimates,
a convergence rate of $k+1$ is observed for $k>2$ \cite{engel:2002},
and a rate of $k$ for polynomial order $k=2$ \cite{wells:2007}.
The example demonstrates that the step from a two-dimensional problem
to a three-dimensional problem is trivial when using the compiler.

\subsection{Evaluating functionals}

\ffc{} can also be used to generate the required code for functionals
(forms of rank zero), which is useful for computing the error when
the exact solution is known or for evaluating various functionals of
the computed solution.  For a problem which has been solved using 5th
order Lagrange basis functions, given the exact solution $u$ and the
finite element solution $u_{h}$, the error $e = u - u_{h}$ and the
\ffc{} input for computing the $L^{2}$ norm of the error is shown in
Table~\ref{tab:L2_norm}. The exact solution has been approximated by
interpolating the exact solution using a continuous 15th order polynomial.
\begin{table}
\caption{Computation of the error in the $L^{2}$ norm (squared).}
\label{tab:L2_norm}
\begin{code}{0.8}
element_u  = FiniteElement("Lagrange", "triangle", 15)
element_uh = FiniteElement("Discontinuous Lagrange", "triangle", 5)

u   = Function(element_u)
u_h = Function(element_uh)

e = u - u_h
M = e*e*dx
\end{code}
\end{table}
The \ffc{} input for computing the mesh-dependent semi-norm of the error
\begin{equation}
  \unorm{e}^{2} =  \int_{\Omega} \nabla e \cdot \nabla e \, \dx
                     + \int_{\Gamma^{0}} \jump{e} \cdot \jump{e} \, \ds,
\end{equation}
is shown in Table~\ref{tab:H1_broken_norm}. Here, $\Gamma^0 = \cup
\partial_i \mathcal{T}$ denotes the union of all interior facets of
the mesh.
\begin{table}
\caption{Computation of the error in a mesh-dependent semi-norm (squared).}
\label{tab:H1_broken_norm}
\begin{code}{0.8}
element_u  = FiniteElement("Lagrange", "triangle", 15)
element_uh = FiniteElement("Discontinuous Lagrange", "triangle", 5)

u   = Function(element_u)
u_h = Function(element_uh)

e = u - u_h
M = dot(grad(e), grad(e))*dx + dot(jump(e), jump(e))*dS
\end{code}
\end{table}

\section{Conclusions}
\label{sec:conclusions}

An approach for automated code generation for discontinuous Galerkin
forms has been presented. The concept is manifest in the form of a
compiler which translates discontinuous Galerkin forms expressed in a
high-level language into efficient low-level code.

A special representation for element tensors for discontinuous Galerkin
methods has also been presented. This representation involves a tensor
contraction and permits the separation of terms which can be computed
\emph{a priori} and terms which are computed at run time. Such a
representation can lead to improved performance relative to conventional
quadrature approaches for a variety of forms. However, the approach is not
generally tractable by hand and necessitates automated code generation
techniques.  The compiler for discontinuous Galerkin forms facilitates
rapid implementation of new and existing approaches and reduces the time
required for code testing through generality, while delivering efficient
code and providing scope for various automated optimisations.
\section*{Acknowledgements}

KB{\O} acknowledges the support of the Netherlands Technology Foundation
STW, the Netherlands Organisation for Scientific Research and the Ministry
of Public Works and Water Management. AL is supported by an Outstanding
Young Investigator grant from the Research Council of Norway, NFR 180450.
\bibliographystyle{siam}
\bibliography{references}
\end{document}

%% file: convergence_biharmonic.tex
\begingroup%
  \makeatletter%
  \newcommand{\GNUPLOTspecial}{%
    \@sanitize\catcode`\%=14\relax\special}%
  \setlength{\unitlength}{0.1bp}%
\begin{picture}(2880,1728)(0,0)%
{\GNUPLOTspecial{"
/gnudict 256 dict def
gnudict begin
/Color false def
/Solid false def
/gnulinewidth 5.000 def
/userlinewidth gnulinewidth def
/vshift -16 def
/dl {10.0 mul} def
/hpt_ 31.5 def
/vpt_ 31.5 def
/hpt hpt_ def
/vpt vpt_ def
/Rounded false def
/M {moveto} bind def
/L {lineto} bind def
/R {rmoveto} bind def
/V {rlineto} bind def
/N {newpath moveto} bind def
/C {setrgbcolor} bind def
/f {rlineto fill} bind def
/vpt2 vpt 2 mul def
/hpt2 hpt 2 mul def
/Lshow { currentpoint stroke M
  0 vshift R show } def
/Rshow { currentpoint stroke M
  dup stringwidth pop neg vshift R show } def
/Cshow { currentpoint stroke M
  dup stringwidth pop -2 div vshift R show } def
/UP { dup vpt_ mul /vpt exch def hpt_ mul /hpt exch def
  /hpt2 hpt 2 mul def /vpt2 vpt 2 mul def } def
/DL { Color {setrgbcolor Solid {pop []} if 0 setdash }
 {pop pop pop 0 setgray Solid {pop []} if 0 setdash} ifelse } def
/BL { stroke userlinewidth 2 mul setlinewidth
      Rounded { 1 setlinejoin 1 setlinecap } if } def
/AL { stroke userlinewidth 2 div setlinewidth
      Rounded { 1 setlinejoin 1 setlinecap } if } def
/UL { dup gnulinewidth mul /userlinewidth exch def
      dup 1 lt {pop 1} if 10 mul /udl exch def } def
/PL { stroke userlinewidth setlinewidth
      Rounded { 1 setlinejoin 1 setlinecap } if } def
/LTw { PL [] 1 setgray } def
/LTb { BL [] 0 0 0 DL } def
/LTa { AL [1 udl mul 2 udl mul] 0 setdash 0 0 0 setrgbcolor } def
/LT0 { PL [] 1 0 0 DL } def
/LT1 { PL [4 dl 2 dl] 0 1 0 DL } def
/LT2 { PL [2 dl 3 dl] 0 0 1 DL } def
/LT3 { PL [1 dl 1.5 dl] 1 0 1 DL } def
/LT4 { PL [5 dl 2 dl 1 dl 2 dl] 0 1 1 DL } def
/LT5 { PL [4 dl 3 dl 1 dl 3 dl] 1 1 0 DL } def
/LT6 { PL [2 dl 2 dl 2 dl 4 dl] 0 0 0 DL } def
/LT7 { PL [2 dl 2 dl 2 dl 2 dl 2 dl 4 dl] 1 0.3 0 DL } def
/LT8 { PL [2 dl 2 dl 2 dl 2 dl 2 dl 2 dl 2 dl 4 dl] 0.5 0.5 0.5 DL } def
/Pnt { stroke [] 0 setdash
   gsave 1 setlinecap M 0 0 V stroke grestore } def
/Dia { stroke [] 0 setdash 2 copy vpt add M
  hpt neg vpt neg V hpt vpt neg V
  hpt vpt V hpt neg vpt V closepath stroke
  Pnt } def
/Pls { stroke [] 0 setdash vpt sub M 0 vpt2 V
  currentpoint stroke M
  hpt neg vpt neg R hpt2 0 V stroke
  } def
/Box { stroke [] 0 setdash 2 copy exch hpt sub exch vpt add M
  0 vpt2 neg V hpt2 0 V 0 vpt2 V
  hpt2 neg 0 V closepath stroke
  Pnt } def
/Crs { stroke [] 0 setdash exch hpt sub exch vpt add M
  hpt2 vpt2 neg V currentpoint stroke M
  hpt2 neg 0 R hpt2 vpt2 V stroke } def
/TriU { stroke [] 0 setdash 2 copy vpt 1.12 mul add M
  hpt neg vpt -1.62 mul V
  hpt 2 mul 0 V
  hpt neg vpt 1.62 mul V closepath stroke
  Pnt  } def
/Star { 2 copy Pls Crs } def
/BoxF { stroke [] 0 setdash exch hpt sub exch vpt add M
  0 vpt2 neg V  hpt2 0 V  0 vpt2 V
  hpt2 neg 0 V  closepath fill } def
/TriUF { stroke [] 0 setdash vpt 1.12 mul add M
  hpt neg vpt -1.62 mul V
  hpt 2 mul 0 V
  hpt neg vpt 1.62 mul V closepath fill } def
/TriD { stroke [] 0 setdash 2 copy vpt 1.12 mul sub M
  hpt neg vpt 1.62 mul V
  hpt 2 mul 0 V
  hpt neg vpt -1.62 mul V closepath stroke
  Pnt  } def
/TriDF { stroke [] 0 setdash vpt 1.12 mul sub M
  hpt neg vpt 1.62 mul V
  hpt 2 mul 0 V
  hpt neg vpt -1.62 mul V closepath fill} def
/DiaF { stroke [] 0 setdash vpt add M
  hpt neg vpt neg V hpt vpt neg V
  hpt vpt V hpt neg vpt V closepath fill } def
/Pent { stroke [] 0 setdash 2 copy gsave
  translate 0 hpt M 4 {72 rotate 0 hpt L} repeat
  closepath stroke grestore Pnt } def
/PentF { stroke [] 0 setdash gsave
  translate 0 hpt M 4 {72 rotate 0 hpt L} repeat
  closepath fill grestore } def
/Circle { stroke [] 0 setdash 2 copy
  hpt 0 360 arc stroke Pnt } def
/CircleF { stroke [] 0 setdash hpt 0 360 arc fill } def
/C0 { BL [] 0 setdash 2 copy moveto vpt 90 450  arc } bind def
/C1 { BL [] 0 setdash 2 copy        moveto
       2 copy  vpt 0 90 arc closepath fill
               vpt 0 360 arc closepath } bind def
/C2 { BL [] 0 setdash 2 copy moveto
       2 copy  vpt 90 180 arc closepath fill
               vpt 0 360 arc closepath } bind def
/C3 { BL [] 0 setdash 2 copy moveto
       2 copy  vpt 0 180 arc closepath fill
               vpt 0 360 arc closepath } bind def
/C4 { BL [] 0 setdash 2 copy moveto
       2 copy  vpt 180 270 arc closepath fill
               vpt 0 360 arc closepath } bind def
/C5 { BL [] 0 setdash 2 copy moveto
       2 copy  vpt 0 90 arc
       2 copy moveto
       2 copy  vpt 180 270 arc closepath fill
               vpt 0 360 arc } bind def
/C6 { BL [] 0 setdash 2 copy moveto
      2 copy  vpt 90 270 arc closepath fill
              vpt 0 360 arc closepath } bind def
/C7 { BL [] 0 setdash 2 copy moveto
      2 copy  vpt 0 270 arc closepath fill
              vpt 0 360 arc closepath } bind def
/C8 { BL [] 0 setdash 2 copy moveto
      2 copy vpt 270 360 arc closepath fill
              vpt 0 360 arc closepath } bind def
/C9 { BL [] 0 setdash 2 copy moveto
      2 copy  vpt 270 450 arc closepath fill
              vpt 0 360 arc closepath } bind def
/C10 { BL [] 0 setdash 2 copy 2 copy moveto vpt 270 360 arc closepath fill
       2 copy moveto
       2 copy vpt 90 180 arc closepath fill
               vpt 0 360 arc closepath } bind def
/C11 { BL [] 0 setdash 2 copy moveto
       2 copy  vpt 0 180 arc closepath fill
       2 copy moveto
       2 copy  vpt 270 360 arc closepath fill
               vpt 0 360 arc closepath } bind def
/C12 { BL [] 0 setdash 2 copy moveto
       2 copy  vpt 180 360 arc closepath fill
               vpt 0 360 arc closepath } bind def
/C13 { BL [] 0 setdash  2 copy moveto
       2 copy  vpt 0 90 arc closepath fill
       2 copy moveto
       2 copy  vpt 180 360 arc closepath fill
               vpt 0 360 arc closepath } bind def
/C14 { BL [] 0 setdash 2 copy moveto
       2 copy  vpt 90 360 arc closepath fill
               vpt 0 360 arc } bind def
/C15 { BL [] 0 setdash 2 copy vpt 0 360 arc closepath fill
               vpt 0 360 arc closepath } bind def
/Rec   { newpath 4 2 roll moveto 1 index 0 rlineto 0 exch rlineto
       neg 0 rlineto closepath } bind def
/Square { dup Rec } bind def
/Bsquare { vpt sub exch vpt sub exch vpt2 Square } bind def
/S0 { BL [] 0 setdash 2 copy moveto 0 vpt rlineto BL Bsquare } bind def
/S1 { BL [] 0 setdash 2 copy vpt Square fill Bsquare } bind def
/S2 { BL [] 0 setdash 2 copy exch vpt sub exch vpt Square fill Bsquare } bind def
/S3 { BL [] 0 setdash 2 copy exch vpt sub exch vpt2 vpt Rec fill Bsquare } bind def
/S4 { BL [] 0 setdash 2 copy exch vpt sub exch vpt sub vpt Square fill Bsquare } bind def
/S5 { BL [] 0 setdash 2 copy 2 copy vpt Square fill
       exch vpt sub exch vpt sub vpt Square fill Bsquare } bind def
/S6 { BL [] 0 setdash 2 copy exch vpt sub exch vpt sub vpt vpt2 Rec fill Bsquare } bind def
/S7 { BL [] 0 setdash 2 copy exch vpt sub exch vpt sub vpt vpt2 Rec fill
       2 copy vpt Square fill
       Bsquare } bind def
/S8 { BL [] 0 setdash 2 copy vpt sub vpt Square fill Bsquare } bind def
/S9 { BL [] 0 setdash 2 copy vpt sub vpt vpt2 Rec fill Bsquare } bind def
/S10 { BL [] 0 setdash 2 copy vpt sub vpt Square fill 2 copy exch vpt sub exch vpt Square fill
       Bsquare } bind def
/S11 { BL [] 0 setdash 2 copy vpt sub vpt Square fill 2 copy exch vpt sub exch vpt2 vpt Rec fill
       Bsquare } bind def
/S12 { BL [] 0 setdash 2 copy exch vpt sub exch vpt sub vpt2 vpt Rec fill Bsquare } bind def
/S13 { BL [] 0 setdash 2 copy exch vpt sub exch vpt sub vpt2 vpt Rec fill
       2 copy vpt Square fill Bsquare } bind def
/S14 { BL [] 0 setdash 2 copy exch vpt sub exch vpt sub vpt2 vpt Rec fill
       2 copy exch vpt sub exch vpt Square fill Bsquare } bind def
/S15 { BL [] 0 setdash 2 copy Bsquare fill Bsquare } bind def
/D0 { gsave translate 45 rotate 0 0 S0 stroke grestore } bind def
/D1 { gsave translate 45 rotate 0 0 S1 stroke grestore } bind def
/D2 { gsave translate 45 rotate 0 0 S2 stroke grestore } bind def
/D3 { gsave translate 45 rotate 0 0 S3 stroke grestore } bind def
/D4 { gsave translate 45 rotate 0 0 S4 stroke grestore } bind def
/D5 { gsave translate 45 rotate 0 0 S5 stroke grestore } bind def
/D6 { gsave translate 45 rotate 0 0 S6 stroke grestore } bind def
/D7 { gsave translate 45 rotate 0 0 S7 stroke grestore } bind def
/D8 { gsave translate 45 rotate 0 0 S8 stroke grestore } bind def
/D9 { gsave translate 45 rotate 0 0 S9 stroke grestore } bind def
/D10 { gsave translate 45 rotate 0 0 S10 stroke grestore } bind def
/D11 { gsave translate 45 rotate 0 0 S11 stroke grestore } bind def
/D12 { gsave translate 45 rotate 0 0 S12 stroke grestore } bind def
/D13 { gsave translate 45 rotate 0 0 S13 stroke grestore } bind def
/D14 { gsave translate 45 rotate 0 0 S14 stroke grestore } bind def
/D15 { gsave translate 45 rotate 0 0 S15 stroke grestore } bind def
/DiaE { stroke [] 0 setdash vpt add M
  hpt neg vpt neg V hpt vpt neg V
  hpt vpt V hpt neg vpt V closepath stroke } def
/BoxE { stroke [] 0 setdash exch hpt sub exch vpt add M
  0 vpt2 neg V hpt2 0 V 0 vpt2 V
  hpt2 neg 0 V closepath stroke } def
/TriUE { stroke [] 0 setdash vpt 1.12 mul add M
  hpt neg vpt -1.62 mul V
  hpt 2 mul 0 V
  hpt neg vpt 1.62 mul V closepath stroke } def
/TriDE { stroke [] 0 setdash vpt 1.12 mul sub M
  hpt neg vpt 1.62 mul V
  hpt 2 mul 0 V
  hpt neg vpt -1.62 mul V closepath stroke } def
/PentE { stroke [] 0 setdash gsave
  translate 0 hpt M 4 {72 rotate 0 hpt L} repeat
  closepath stroke grestore } def
/CircE { stroke [] 0 setdash 
  hpt 0 360 arc stroke } def
/Opaque { gsave closepath 1 setgray fill grestore 0 setgray closepath } def
/DiaW { stroke [] 0 setdash vpt add M
  hpt neg vpt neg V hpt vpt neg V
  hpt vpt V hpt neg vpt V Opaque stroke } def
/BoxW { stroke [] 0 setdash exch hpt sub exch vpt add M
  0 vpt2 neg V hpt2 0 V 0 vpt2 V
  hpt2 neg 0 V Opaque stroke } def
/TriUW { stroke [] 0 setdash vpt 1.12 mul add M
  hpt neg vpt -1.62 mul V
  hpt 2 mul 0 V
  hpt neg vpt 1.62 mul V Opaque stroke } def
/TriDW { stroke [] 0 setdash vpt 1.12 mul sub M
  hpt neg vpt 1.62 mul V
  hpt 2 mul 0 V
  hpt neg vpt -1.62 mul V Opaque stroke } def
/PentW { stroke [] 0 setdash gsave
  translate 0 hpt M 4 {72 rotate 0 hpt L} repeat
  Opaque stroke grestore } def
/CircW { stroke [] 0 setdash 
  hpt 0 360 arc Opaque stroke } def
/BoxFill { gsave Rec 1 setgray fill grestore } def
/BoxColFill {
  gsave Rec
  /Fillden exch def
  currentrgbcolor
  /ColB exch def /ColG exch def /ColR exch def
  /ColR ColR Fillden mul Fillden sub 1 add def
  /ColG ColG Fillden mul Fillden sub 1 add def
  /ColB ColB Fillden mul Fillden sub 1 add def
  ColR ColG ColB setrgbcolor
  fill grestore } def
%
%
/PatternFill { gsave /PFa [ 9 2 roll ] def
    PFa 0 get PFa 2 get 2 div add PFa 1 get PFa 3 get 2 div add translate
    PFa 2 get -2 div PFa 3 get -2 div PFa 2 get PFa 3 get Rec
    gsave 1 setgray fill grestore clip
    currentlinewidth 0.5 mul setlinewidth
    /PFs PFa 2 get dup mul PFa 3 get dup mul add sqrt def
    0 0 M PFa 5 get rotate PFs -2 div dup translate
	0 1 PFs PFa 4 get div 1 add floor cvi
	{ PFa 4 get mul 0 M 0 PFs V } for
    0 PFa 6 get ne {
	0 1 PFs PFa 4 get div 1 add floor cvi
	{ PFa 4 get mul 0 2 1 roll M PFs 0 V } for
    } if
    stroke grestore } def
/Symbol-Oblique /Symbol findfont [1 0 .167 1 0 0] makefont
dup length dict begin {1 index /FID eq {pop pop} {def} ifelse} forall
currentdict end definefont pop
end
gnudict begin
gsave
0 0 translate
0.100 0.100 scale
0 setgray
newpath
1.000 UL
LTb
225 100 M
63 0 V
2517 0 R
-63 0 V
1.000 UL
LTb
225 168 M
31 0 V
2549 0 R
-31 0 V
225 258 M
31 0 V
2549 0 R
-31 0 V
225 304 M
31 0 V
2549 0 R
-31 0 V
225 325 M
63 0 V
2517 0 R
-63 0 V
1.000 UL
LTb
225 393 M
31 0 V
2549 0 R
-31 0 V
225 483 M
31 0 V
2549 0 R
-31 0 V
225 529 M
31 0 V
2549 0 R
-31 0 V
225 551 M
63 0 V
2517 0 R
-63 0 V
1.000 UL
LTb
225 619 M
31 0 V
2549 0 R
-31 0 V
225 708 M
31 0 V
2549 0 R
-31 0 V
225 754 M
31 0 V
2549 0 R
-31 0 V
225 776 M
63 0 V
2517 0 R
-63 0 V
1.000 UL
LTb
225 844 M
31 0 V
2549 0 R
-31 0 V
225 934 M
31 0 V
2549 0 R
-31 0 V
225 980 M
31 0 V
2549 0 R
-31 0 V
225 1002 M
63 0 V
2517 0 R
-63 0 V
1.000 UL
LTb
225 1070 M
31 0 V
2549 0 R
-31 0 V
225 1159 M
31 0 V
2549 0 R
-31 0 V
225 1205 M
31 0 V
2549 0 R
-31 0 V
225 1227 M
63 0 V
2517 0 R
-63 0 V
1.000 UL
LTb
225 1295 M
31 0 V
2549 0 R
-31 0 V
225 1385 M
31 0 V
2549 0 R
-31 0 V
225 1431 M
31 0 V
2549 0 R
-31 0 V
225 1453 M
63 0 V
2517 0 R
-63 0 V
1.000 UL
LTb
225 1520 M
31 0 V
2549 0 R
-31 0 V
225 1610 M
31 0 V
2549 0 R
-31 0 V
225 1656 M
31 0 V
2549 0 R
-31 0 V
225 1678 M
63 0 V
2517 0 R
-63 0 V
1.000 UL
LTb
2805 100 M
0 31 V
0 1547 R
0 -31 V
2685 100 M
0 31 V
0 1547 R
0 -31 V
2577 100 M
0 63 V
0 1515 R
0 -63 V
1.000 UL
LTb
1869 100 M
0 31 V
0 1547 R
0 -31 V
1455 100 M
0 31 V
0 1547 R
0 -31 V
1161 100 M
0 31 V
0 1547 R
0 -31 V
933 100 M
0 31 V
0 1547 R
0 -31 V
747 100 M
0 31 V
0 1547 R
0 -31 V
589 100 M
0 31 V
0 1547 R
0 -31 V
453 100 M
0 31 V
0 1547 R
0 -31 V
333 100 M
0 31 V
0 1547 R
0 -31 V
225 100 M
0 63 V
0 1515 R
0 -63 V
1.000 UL
LTb
1.000 UL
LTb
225 100 M
2580 0 V
0 1578 V
-2580 0 V
225 100 L
0.750 UP
LTb
LTb
LTb
LTb
LTb
LTb
LTb
LTb
0.750 UP
1.000 UL
LTb
1455 1113 M
0 -131 V
681 0 V
-681 131 V
1455 1113 Pnt
1455 982 Pnt
2136 982 Pnt
1455 1113 Pnt
0.750 UP
1.000 UL
LTb
1041 893 M
0 -206 V
537 0 V
1041 893 L
1041 893 Pnt
1041 687 Pnt
1578 687 Pnt
1041 893 Pnt
0.750 UP
1.000 UL
LTb
747 704 M
0 -389 V
810 0 V
747 704 L
747 704 Pnt
747 315 Pnt
1557 315 Pnt
747 704 Pnt
0.750 UP
1.000 UL
LT0
LTb
LT0
2608 1571 M
147 0 V
372 1430 M
708 -204 V
708 -128 V
2496 968 L
372 1430 Pls
1080 1226 Pls
1788 1098 Pls
2496 968 Pls
2681 1571 Pls
0.750 UP
1.000 UL
LT1
LTb
LT1
2608 1483 M
147 0 V
372 1321 M
1080 915 L
1788 654 L
372 1321 Crs
1080 915 Crs
1788 654 Crs
2681 1483 Crs
0.750 UP
1.000 UL
LT2
LTb
LT2
2608 1395 M
147 0 V
372 933 M
1080 585 L
1788 218 L
372 933 Star
1080 585 Star
1788 218 Star
2681 1395 Star
1.000 UL
LTb
225 100 M
2580 0 V
0 1578 V
-2580 0 V
225 100 L
0.750 UP
stroke
grestore
end
showpage
}}%
\fontsize{10}{\baselineskip}\selectfont
\put(2583,1395){\makebox(0,0)[r]{$k = 4$}}%
\put(2583,1483){\makebox(0,0)[r]{$k = 3$}}%
\put(2583,1571){\makebox(0,0)[r]{$k = 2$}}%
\put(676,498){\makebox(0,0)[l]{5}}%
\put(1117,259){\makebox(0,0)[l]{1}}%
\put(970,778){\makebox(0,0)[l]{4}}%
\put(1274,630){\makebox(0,0)[l]{1}}%
\put(1384,1036){\makebox(0,0)[l]{2}}%
\put(1760,926){\makebox(0,0)[l]{1}}%
\put(-80,658){%
\special{ps: gsave currentpoint currentpoint translate
270 rotate neg exch neg exch translate}%
\makebox(0,0)[lb]{\shortstack{$||u - u_h||$}}%
\special{ps: currentpoint grestore moveto}%
}%
\put(1455,-34){\makebox(0,0)[l]{$h$}}%
\put(225,50){\makebox(0,0){ 1}}%
\put(2577,50){\makebox(0,0){ 0.1}}%
\put(200,1678){\makebox(0,0)[r]{ 10}}%
\put(200,1453){\makebox(0,0)[r]{ 1}}%
\put(200,1227){\makebox(0,0)[r]{ 0.1}}%
\put(200,1002){\makebox(0,0)[r]{ 0.01}}%
\put(200,776){\makebox(0,0)[r]{ 0.001}}%
\put(200,551){\makebox(0,0)[r]{ 1e-04}}%
\put(200,325){\makebox(0,0)[r]{ 1e-05}}%
\put(200,100){\makebox(0,0)[r]{ 1e-06}}%
\end{picture}%
\endgroup
 

%% file: paper.bbl
\begin{thebibliography}{10}

\bibitem{logg:manual:04}
{\sc M.~S. Aln\ae{}s, A.~Logg, K.-A. Mardal, O.~Skavhaug, and H.-P Langtangen},
  {\em UFC Specification and User Manual}, 2008.
\newblock URL: \url{http://www.fenicsproject.org/}.

\bibitem{AlnLog2008}
{\sc Martin~Sandve Aln{\ae}s, Anders Logg, Kent-Andre Mardal, Ola Skavhaug, and
  Hans~Petter Langtangen}, {\em Unified framework for finite element assembly},
  International Journal of Computational Science and Engineering, 4 (2009),
  pp.~231--244.

\bibitem{www:SyFi}
{\sc M.~S. Aln\ae{}s and K.-A. Mardal}, {\em {S}y{F}i: Symbolic Finite
  Elements}, 2008.
\newblock \emph{http://www.fenicsproject.org/}.

\bibitem{arnold_et_al:2002}
{\sc D.~N. Arnold, F.~Brezzi, B.~Cockburn, and L.~D. Marini}, {\em Unified
  analysis for discontinuous {G}alerkin methods for elliptic problems}, SIAM
  Journal on Numerical Analysis, 39 (2002), pp.~1749--1779.

\bibitem{baker:1990}
{\sc G.~A. Baker, W.~N. Jureidini, and O.~A. Karakashian}, {\em Piecewise
  solenoidal vector fields and the {S}tokes problem}, SIAM Journal on Numerical
  Analysis, 27 (1990), pp.~1466--1485.

\bibitem{engel:2002}
{\sc G.~Engel, K.~Garikipati, T.~J.~R. Hughes, M.~G. Larson, and R.~L. Taylor},
  {\em Continuous/discontinuous finite element approximations of fourth-order
  elliptic problems in structural and continuum mechanics with applications to
  thin beams and plates, and strain gradient elasticity}, Computer Methods in
  Applied Mechanics and Engineering, 191 (2002), pp.~3669--3750.

\bibitem{Hug87}
{\sc T.~J.~R. Hughes}, {\em The Finite Element Method: Linear Static and
  Dynamic Finite Element Analysis}, Dover Inc., 2000.

\bibitem{hughes:2006}
{\sc T.~J.~R. Hughes, G.~Scovazzi, P.~B. Bochev, and A.~Buffa}, {\em A
  multiscale discontinuous {G}alerkin method with the computational structure
  of a continuous {G}alerkin method}, Computer Methods in Applied Mechanics and
  Engineering, 195 (2006), pp.~2761--2787.

\bibitem{Kir04}
{\sc R.~C. Kirby}, {\em {FIAT}: A new paradigm for computing finite element
  basis functions}, ACM Transactions on Mathematical Software (TOMS), 30
  (2004), pp.~502--516.

\bibitem{Kir05}
\leavevmode\vrule height 2pt depth -1.6pt width 23pt, {\em Optimizing {FIAT}
  with {Level 3 BLAS}}, ACM Transactions on Mathematical Software (TOMS), 32
  (2006), pp.~223--235.

\bibitem{logg:article:07}
{\sc R.~C. Kirby, M.~G. Knepley, A.~Logg, and L.~R. Scott}, {\em Optimizing the
  evaluation of finite element matrices}, SIAM Journal on Scientific Computing,
  27 (2005), pp.~741--758.

\bibitem{logg:article:10}
{\sc R.~C. Kirby and A.~Logg}, {\em A compiler for variational forms}, ACM
  Transactions on Mathematical Software (TOMS), 32 (2006), pp.~417--444.

\bibitem{logg:article:11}
\leavevmode\vrule height 2pt depth -1.6pt width 23pt, {\em Efficient
  compilation of a class of variational forms}, ACM Transactions on
  Mathematical Software (TOMS), 33 (2007).

\bibitem{KirLog2008a}
{\sc R.~C. Kirby and A.~Logg}, {\em Benchmarking domain-specific compiler
  optimizations for variational forms}, ACM Transactions on Mathematical
  Software, 35 (2008), pp.~1--18.

\bibitem{logg:article:09}
{\sc R.~C. Kirby, A.~Logg, L.~R. Scott, and A.~R. Terrel}, {\em Topological
  optimization of the evaluation of finite element matrices}, SIAM Journal on
  Scientific Computing, 28 (2006), pp.~224--240.

\bibitem{Labeur:2007}
{\sc R.~J. Labeur and G.~N. Wells}, {\em A {G}alerkin interface stabilisation
  method for the advection-diffusion and incompressible {N}avier-{S}tokes
  equations}, Computer Methods in Applied Mechanics and Engineering, 196
  (2007), pp.~4985--5000.

\bibitem{Lan99}
{\sc H.~P. Langtangen}, {\em Computational Partial Differential Equations --
  Numerical Methods and Diffpack Programming}, Lecture Notes in Computational
  Science and Engineering, Springer, 1999.

\bibitem{logg:article:12}
{\sc A.~Logg}, {\em Automating the finite element method}, Archives of
  Computational Methods in Engineering, 14 (2007), pp.~93--138.

\bibitem{logg:manual:02}
\leavevmode\vrule height 2pt depth -1.6pt width 23pt, {\em {FFC} User Manual},
  2007.
\newblock URL: \url{http://www.fenics.org/ffc/}.

\bibitem{www:dolfin}
{\sc A.~Logg, G.~N. Wells, J.~Hoffman, J.~Jansson, et~al.}, {\em {DOLFIN}: A
  general-purpose finite element library}.
\newblock \emph{http://www.fenicsproject.org/}.

\bibitem{Mardal.2006.6}
{\sc K.-A. Mardal}, {\em {SyFi} - an element matrix factory}, in Proceedings of
  the 8th international conference on Applied parallel computing: state of the
  art in scientific computing, PARA'06, Springer-Verlag, Berlin, Heidelberg,
  2007, pp.~703--711.

\bibitem{wells:2007}
{\sc G.~N. Wells and N.~T. Dung}, {\em A {$C^{0}$} discontinuous {G}alerkin
  formulation for {K}irchhoff plates}, Computer Methods in Applied Mechanics
  and Engineering, 196 (2007), pp.~3370--3380.

\bibitem{fenics}
{\em The {FE}ni{CS} {P}roject}.
\newblock \url{http://www.fenicsproject.org}.

\bibitem{ZieTay67}
{\sc O.~C. Zienkiewicz, R.~L. Taylor, and J.~Z. Zhu}, {\em The Finite Element
  Method: Its Basis and Fundamentals}, Butterworth-Heinemann, sixth~ed., 2005.

\end{thebibliography}
